\documentclass[pdflatex,sn-mathphys-num,iicol]{sn-jnl}% Math and Physical
\usepackage{graphicx}%
\usepackage{multirow}%
\usepackage{amsmath,amssymb,amsfonts}%
\usepackage{amsthm}%
\usepackage{mathrsfs}%
\usepackage[title]{appendix}%
\usepackage{xcolor}%
\usepackage{textcomp}%
\usepackage{manyfoot}%
\usepackage{booktabs}%
\usepackage{algorithm}%
\usepackage{algorithmicx}%
\usepackage{algpseudocode}%
\usepackage{listings}%
\usepackage{siunitx}%
\usepackage{hyperref}
\usepackage{upgreek}
\theoremstyle{thmstyleone}%
\theoremstyle{thmstyletwo}%

\theoremstyle{thmstylethree}%

\begin{document}

\title[Piezoelectricity: A Brief History of its Discovery and
Physical Principles Using the Example of
Low Quartz]{Piezoelectricity: A Brief History of its Discovery and
Physical Principles Using the Example of
Low Quartz}


\author*[1]{\fnm{Rüdiger
G.}\sur{Ballas}}\email{ruediger.ballas@wb-fernstudium.de}

\author[1]{\fnm{Nataliya} \sur{Koev}}\email{nataliya.koev@wb-fernstudium.de}


\affil*[1]{\orgdiv{Department of Engineering Sciences}, \orgname{Wilhelm
Büchner Hochschule},
\orgaddress{\street{Hilpertstrasse 31}, \city{Darmstadt}, \postcode{64295},
\state{Hesse},
\country{Germany}}}

%\affil[2]{\orgdiv{Department}, \orgname{Organization},
%\orgaddress{\street{Street}, \city{City}, \postcode{10587}, \state{State},
%\country{Country}}}
%
%\affil[3]{\orgdiv{Department}, \orgname{Organization},
%\orgaddress{\street{Street}, \city{City}, \postcode{610101}, \state{State},
%\country{Country}}}


\abstract{Piezoelectricity is a key phenomenon in
solid-state physics
and has 
significant
technological relevance. In teaching, however, it is often treated either
without a
historical context or an explicit link to the microscopic
crystal structure, meaning that a crucial level of understanding
remains untapped. This study combines the
history of the discovery of piezoelectricity with its physical
description within a didactically coherent framework. Building on the work
of the brothers, Jacques and Pierre Curie (1880–1882), it demonstrated how
investigations into pyroelectricity led to the identification of polarization
induced by mechanical compression in hemihedral crystals such as tourmaline
and quartz. The development from early qualitative experiments towards the
first quantitative determination of piezoelectric constants is
traced in a structured manner. Using low quartz as an example, the
microscopic
cause of piezoelectricity was attributed to the asymmetric displacements of
silicon and oxygen ions
within the orthosilicate ion tetrahedral network. This 
leads to the formation of macroscopic polarization owing to
uncompensated dipole moments in the crystal lattice. The direct and reciprocal
piezoelectric effects are qualitatively distinguished and
discussed
in
the
context of
mechanical-electrical coupling. The article is aimed at
students in physics and engineering as well as lecturers. It was
conceived as a didactic review article (tutorial article) and serves as a
structured introduction to the physical and historical foundations of
piezoelectricity.}




\maketitle

\section{Introduction}

Piezoelectricity is one of the fundamental electromechanical
phenomena and has played a central role in the development of modern
sensor, actuator, and transducer technologies. Since its discovery by
Jacques and Pierre Curie in 1880, it has attracted sustained interest
both as a physical effect and as a basis for numerous technical
applications. Despite its scientific and technological importance,
specialist literature rarely provides a concise, historically
contextualized account of the underlying physical principles. This
paper is aimed at physics and engineering students, as well as
specialists in related disciplines, and offers a structured overview of
the subject area. It pursues a two-fold objective: in the historical
section (Section~\ref{sec:Entdeckung}) the development from the
experimental discovery of piezoelectricity in the late 19th century is traced,
whereas in the physical section (Section~\ref{sec:Physik}) the
microscopic and macroscopic fundamentals are systematically derived using the
example of low quartz. This work is intended as a supplementary
introduction to the standard literature on piezoelectricity.

\section{The discovery of piezoelectricity}
\label{sec:Entdeckung}

\begin{quote}
Crystals that have one or more axes with different ends, that is hemihedral
(semi-symmetrical) crystals with oblique faces, possess a particular physical
property: when the temperature changes, two electric poles with opposite signs
are generated at the ends of these axes. This phenomenon is known as
pyroelectricity. We discovered a new method for generating polar
electricity in the same crystals in which they are subjected to pressure
fluctuations along their hemihedral axes~\cite{Curie1880}.
\end{quote}

On August 2, 1880, brothers Jacques and Pierre Curie announced the
discovery of the piezoelectric effect of the French Academy of Sciences. As
early as the 8th of April of the same year, Jacques Curie informed the
French Mineralogical Society that, together with Pierre, he discovered that
the compression of asymmetric crystals along their hemihedral axes induces
electrical polarization. When these crystals were
decompressed in the same
direction, an electrical effect of opposite sign was produced. On the other, 
amorphous
materials showed no electrical effects under
pressure~\cite{Katzir2006}.

The Curies compared an unknown phenomenon and its properties with the known
phenomenon of pyroelectricity, which occurs in the same crystals. In their
investigation of six types of crystals (tourmaline, zinc blende, boracite,
topaz, calamine, and quartz), the brothers found that in all cases, the
electrical effect of compression corresponds to that of cooling, and that of
decompression corresponds to that of heating, with regard to the directions
and signs of the generated charge. However, they found no correlation between
the magnitude
of the effects of heating and pressure. They noted that the elastic constants
of these crystals are positive (except for calamine,
whose coefficient is unknown)~\cite{Curie1880}. Thus, heating of these crystals
causes thermal expansion, whereas decompression causes mechanical expansion of
the same type. The correspondence between the phenomena of pressure and
temperature change led them to regard both as manifestations of the effects of
contraction and expansion~\cite{Katzir2006}.

\begin{quote}
Whatever the underlying cause may be (as they wrote), whenever a
non-conductive hemihedral crystal with an inclined face contracts, electric
poles are formed in a specific direction; whenever the crystal expands,
electricity is released in the opposite direction~\cite{Curie1880}.
\end{quote}

\begin{figure}[b]
\centering
\includegraphics[width=\linewidth]
{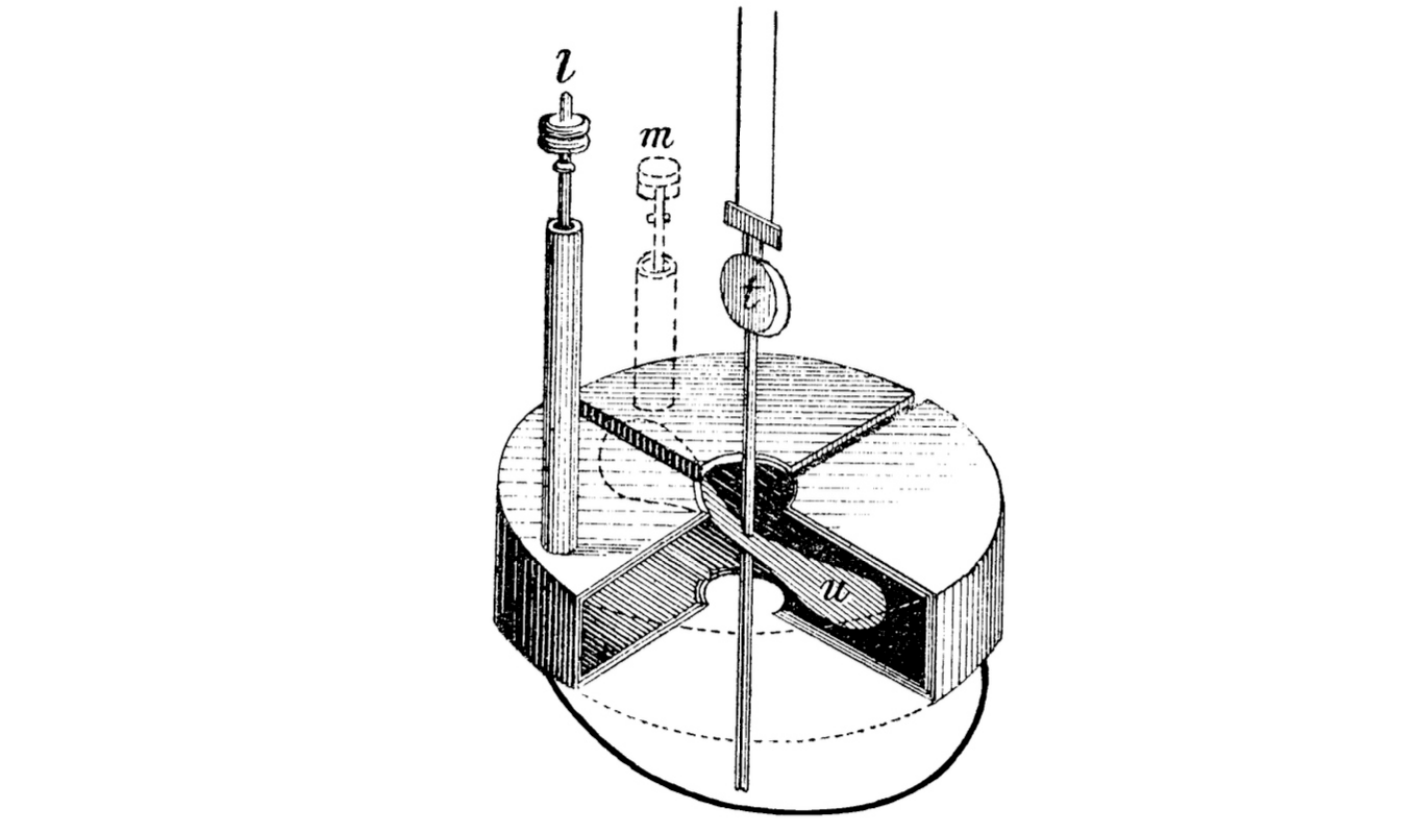}
\caption{Thomson quadrant electrometer: The two opposite square sections are
electrically connected to one another and insulated from the adjacent squares.
They are usually connected via $l$ and $m$ to two different parts of the
circuit, whilst the needle $u$ is energised by a third source. The needle is
suspended without electrical contact with the sections and is rotated by the
electrical force exerted by the quadrants against the torsion of a platinum
wire. Its deflection is observed by the movement of $t$ and is approximately
proportional to the voltage difference between the
sectors (reproduced from~\cite{Graetz1905}).}
\label{fig:Thomson}
\end{figure}

The experiments performed to obtain these results were
relatively straightforward. A piece of the crystal was placed perpendicular to
its half-axis between the two copper plates. These plates, which were
electrically insulated from the surrounding environment, were clamped in a vice
to compress and 
release the sample. The Curies connected the plates in two different
configurations to a Thomson quadrant electrometer, which was probably the most
common device for measuring electrical voltage at that time (see
Figure~\ref{fig:Thomson})~\cite{Graetz1905}.
The instrument consisted of a cylindrical brass container divided
(electrically) into four quadrants; the opposite quadrants are connected by a
wire. An aluminum strip (known as a pointer) inside the container moves freely
under the electrical influence of quadrants, resisting torsion, with its
position visible from the outside.

In the first configuration, the Curies connected two copper plates to
two pairs of opposite quadrants and electrified the needle. This
corresponded to the standard use of the Thomson electrometer, which indicated
the voltage difference (on a relative scale) between adjacent quadrants, and
thus between the copper plates at the edges of the crystal. In an alternative
arrangement, ``the electricity of each plate could be measured individually;
to do this, it was sufficient to connect one of the copper plates to earth and
the other to the electrometer’s needle, while the two pairs of sectors were
charged by a battery''~\cite{Curie1880}. However, this configuration is
unsuitable for quantitative measurements. Initially, the Curies
focused on
qualitative questions. These experiments were specifically designed
to investigate the electrical effect of pressure on hemihedral crystals. No
accidents have been discovered~\cite{Katzir2006}.

The measurement of electrical voltage was carried out using the methods common
at the time. In particular, it resembled the technique that Charles Friedel had
used a year earlier in his investigations into pyroelectricity, and which is
mentioned by the Curies. Jacques Curie was Friedel’s assistant in the
mineralogical laboratory of the Faculty of Natural Sciences at the Sorbonne in
Paris. The challenge in these experiments lay in cutting the crystals along
their crystal axes, whereby two parallel faces had to be arranged perpendicular
to the axis under investigation. However, such precisely cut crystal prisms
were available from manufacturers and were used in several mineralogical
laboratories, including that of the Curies. They therefore presumably used
ready-made prisms and did not have to cut the crystals themselves. The
remaining components of their apparatus (in this and later experiments)
consisted of standard laboratory equipment~\cite{Katzir2006}.

The brothers conducted their first experiments on piezoelectricity in
Friedel’s laboratory. Like his brother, Pierre was an assistant in the Faculty
of Natural Sciences in Paris. That winter, Pierre worked with Paul Desains in
the latter’s physics laboratory on the wavelength of ``heat waves'',
which were soon referred to as thermal radiation. In June, they submitted a
joint paper on this subject to the Academy of Sciences~\cite{Katzir2006}.

However, Jacques and Pierre Curie’s work on piezoelectricity had already begun,
as reported in April 1880. Thus, until summer, Pierre worked
simultaneously with Desains on thermal radiation and his brother on
piezoelectricity. It is likely that Jacques and Pierre Curie returned to their
studies on piezoelectricity with full force in summer and submitted two
papers to the Academy of Sciences in August~\cite{Katzir2006}.

In these publications, the Curies reported on the piezoelectricity of four
further crystal types known to be pyroelectric, which had not been mentioned in
their April publication~\cite{Cady1946}. Based on the ten crystal
types
studied, they drew general conclusions regarding the relationship between the
crystallographic structure and the generation of electricity through changes in
pressure or temperature. They formulated a rule that Haüy had already
established for pyroelectricity and concluded that, upon contraction, the
positive pole is always located at the end of the axis, where the angle between
the axis and the crystal face is sharper. By comparing the crystal axes excited
by pressure with the unexcited axes, they developed rules governing
the electrical effect. In 1882, they clearly summarized these rules based on
crystal symmetry~\cite{Katzir2006}.

To exhibit the properties of an electric axis in a crystal,
the crystal must lack the same symmetry element that is also absent in an
electric field directed along that direction, that is,
\begin{enumerate}
	\item that it (the crystal) has no centre;
	\item that it has no plane of symmetry perpendicular to the direction in
question;
	\item that it has no axis of symmetry of even order perpendicular to this
direction.
\end{enumerate}
These conditions are necessary, and experiments show that they are
sufficient for crystals~\cite{Curie1882}.

In the six months following their discovery, Jacques and Pierre Curie continued
to investigate the properties of this newly discovered phenomenon. Their
initial
qualitative experiments confirmed the existence of this phenomenon, drawing
parallels with pyroelectricity and demonstrating its connection to crystal
symmetry. The Curies did not pursue further research into the relationship
between piezoelectricity and crystal structure; this was later taken up by 
Wilhelm Gottlieb Hankel, Wilhelm Conrad Röntgen, and others. Instead, they
concentrated on systematic quantitative experiments to determine the laws
governing the generation of charges through pressure. What motivated them to do
this? First, Gaugain formulated comparable laws for pyroelectricity 25
years earlier. The close connection they recognized between the two phenomena
suggests that the laws governing electrical generation must be similar in both
cases. Second, in the late 19th century, quantitative laws were the preferred
method for describing physical relationships. While quantitative laws for
pyroelectricity were not formulated until a century after its discovery, this
was achieved for piezoelectricity just one year after its discovery, a sign
of
the advancing quantification of electricity research in the 19th century.
Third, such rules were able to support the Curies' explanatory models of the
phenomenon~\cite{Katzir2006}.

After Jean-Monthé Gaugain left the École Polytechnique in 1830 for political
reasons and managed various metallurgical works, he returned to Paris in
1851 to devote himself to the study and teaching of electricity without
holding a permanent post at a university. His early work focused on the
generation of electricity and its interactions with other substances, which
led him to investigate pyroelectricity~\cite{Poggendorff1863}. In
1856, he conducted ``numerous experiments'' on more than 30 tourmaline
samples under various heating and cooling conditions. These experiments
surpassed all previous ones in terms of precision and rigor with regard to
experimental error. These were the first comprehensive quantitative
measurements aimed at formulating general rules regarding the dependence of the
magnitude of the effect on external factors, such as temperature changes and
crystal
dimensions. However, Gaugain did not measure absolute values but only relative
values, which were sufficient for his purposes. Based on this data, he
developed three empirical laws~\cite{Katzir2006}:

\begin{enumerate}
\item The amount of electricity generated by a single prism is proportional to
its cross-sectional area and independent of its length.
\item The amount of electricity generated by
tourmaline when the
temperature decreases by
a
certain number of degrees is independent of the cooling duration.
\item The amount of electricity generated by the tourmaline when the
temperature
rises by a
certain number of degrees corresponds exactly to the amount of electricity
generated when
the temperature falls by the same amount.
\end{enumerate}
Points 1) and 2) indicate that the amount of heat generated remains constant
for
every change in temperature, regardless of the absolute temperature and 
total change in temperature~\cite{Gaugain1859}.

In 1828, Antoine César Becquerel performed the first quantitative measurements
of electricity on a cooled specimen using an electrometer to support his
theories on pyroelectricity. These early quantitative investigations by
Becquerel and, shortly afterwards, by James Forbes made it possible to analyze
the dependence of the phenomenon on the dimensions of the test specimens.
Becquerel observed that long tourmaline crystals showed no electrical
effect, an observation that Forbes doubts. In his experiments, Forbes noted
differences in the intensity of the effect (the electric charge) between
longer and shorter specimens but attributed this primarily to internal
differences within the specimens. When the specimen was divided in a 1:3 ratio,
it became apparent that both parts exhibited a comparable intensity, from which
he concluded that the length had no influence on the intensity. On the other
hand, the surface area of the crystal significantly influences the
intensity: the larger the area, the stronger the effect~\cite{Forbes1834}.
However, Forbes were unable to establish a mathematical relationship between
the surface area and the electrical effect, something Gaugain succeeded in
doing two decades later~\cite{Katzir2006}.

Similar to Gaugain, the Curies focused on a single type of crystal --
tourmaline. The choice of tourmaline seemed obvious, as it is known for its
pronounced pyroelectric and piezoelectric properties and is regarded as the
standard crystal in pyroelectric research. In their efforts to establish
quantitative laws, their new experiments surpassed the precision of earlier
investigations. The pressure was generated by a wooden lever (presumably
weighted on the opposite side), rather than vice versa. One of the two
copper plates surrounding the crystal was coupled to the needle of a
Thomson electrometer, whereas the other was earthed. By earthing the plate,
they followed Gaugain, who found that this increased the deflection of the
electroscope during pyroelectric measurements~\cite{Gaugain1859}. Each
time, they placed a tourmaline prism between the (insulated) copper plates,
aligned perpendicular to its principal axis. They used prisms of varying
lengths between \SI{0.5}{\milli\metre} and \SI{15}{\milli\metre}
with a constant surface area, and surfaces ranging from \SI{2}{\milli\metre^2}
to \SI{2}{\centi\metre^2}. Using different weights, they collected sufficient
experimental data to define the fundamental principles of electricity
generation by pressure in tourmaline~\cite{Katzir2006}, \cite{Curie1881a}.

Jacques and Pierre Curie did not publish detailed findings of their
investigations. Instead, in January 1881, they submitted their findings to the
Academy of Sciences. They wrote the following five points~\cite{Katzir2006}:

\begin{enumerate}
\item The two ends of the tourmaline release equal amounts of electricity with
opposite signs.
\item The amount released by a given increase in pressure has the opposite sign
and corresponds to the amount produced by an equally large decrease in
pressure.
\item This quantity is proportional to the change in pressure.
\item It does not depend on the length of the tourmaline.
\item For the same pressure variation (sic) per unit area, it is proportional
to the area.
\end{enumerate}

An important insight can be derived from the last two laws, namely that ``for a
constant change in pressure, the amount of electricity released is independent
of the dimensions of the tourmaline''~\cite{Curie1881a}. These laws relate to
the ``amount of electricity'', that is the electric charge generated by a
change in pressure on the crystal surface and measured by the electrometer.
Because the laws are merely relative, the absolute magnitude of the charge is
irrelevant, which is why one can rely on the voltage measurement by the
electrometer, which is proportional to the charge (at constant capacitance),
since ``the capacitance of the copper plates \ldots was always negligible
compared to the capacitance of the electrometer''~\cite{Curie1881a}. Later, 
physicists preferred to refer to electrical polarization (dipole moment
density, also known as electric moment), an inherent property of the crystal.
By ``pressure'' the Curies meant the total pressure, i.e. the force or weight
in their experiment, not the force per unit area. It is easy to demonstrate
that their conclusion also holds for voltage and polarization, instead of
weight and charge. The brothers concluded that these laws were consistent with
those of pyroelectricity. This correspondence is explained by their
hypothesis that ``the contraction or expansion along the tourmaline axis''
causes both effects. Previously, they suggested that the phenomena arise from
contraction and expansion; after confirming the qualitative rules, they made
this causal relationship explicit~\cite{Curie1881a}. Shortly afterwards, they
examined quartz and found that its electrification by pressure follows the same
rules~\cite{Katzir2006}, \cite{Curie1882}.

According to the empirical laws developed by the Curies, the electrical effect
of a change in pressure on the crystals is linearly dependent. Therefore, each
crystal has a characteristic coefficient (or several coefficients) that
indicates the amount of electrical charge generated by an increase or decrease
in pressure. Having confirmed the accuracy of this relationship, the brothers
conducted an experiment in the first half of 1881 to determine the tourmaline 
and quartz coefficients. This marked a further advance in the quantification of
the phenomena, surpassing Gaugain’s pyroelectric measurements, which provided
only relationships, but no numerical values. In their previous experiments,
they merely determined the electrical voltage, which they knew was proportional
to the charge, but did not know the proportionality factor (which represents
the capacitance of the system). To determine the values of the coefficients, it
is necessary to establish a precise quantitative relationship between the
electrical voltage and charge~\cite{Katzir2006}.

\begin{figure}[b]
\centering
\includegraphics[width=\linewidth]
{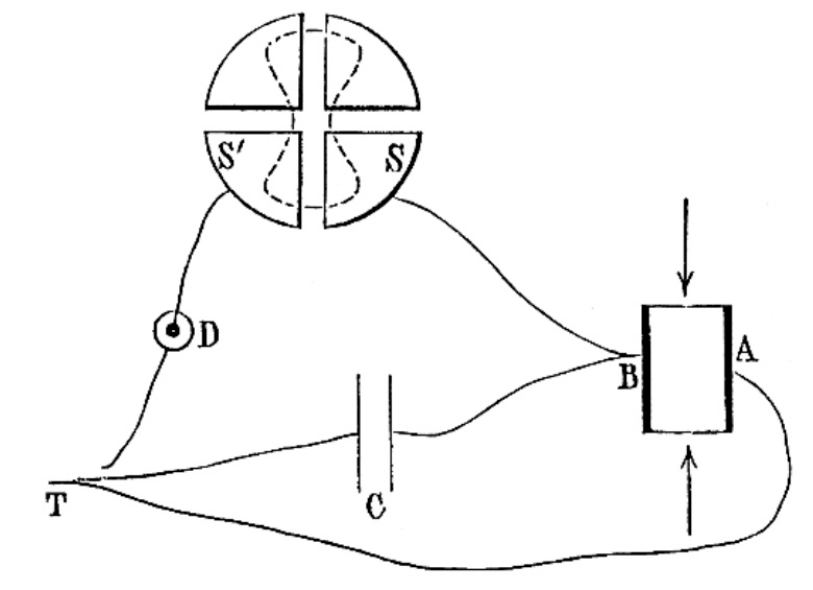}
\caption{Curie’s experimental measurement of the piezoelectric
constants (reproduced from~\cite{Curie1882}).}
\label{fig:Messaufbau}
\end{figure}

To this end, the Curies designed a new experimental apparatus based on the one
they had previously used (see Figure~\ref{fig:Messaufbau})~\cite{Curie1882}).
A crystal rod was placed between two copper plates perpendicular along the
hemihedral (semi-symmetrical) axis. One plate (A) was connected to earth
potential, whilst the other (B) was coupled to a cylindrical capacitor (C) and
a sector of the Thomson electrometer, whose second sector was connected to a
``Daniell cell'' (D) -- a battery with a known and constant potential of
\SI{1.2}{\volt} (whilst the needle is charged). The other pole is then earthed.
They placed known weights directly onto the crystal via a support (potence) and
exchanged standardized cylindrical capacitors until the charged needle of the
electrometer stabilized at its zero point between the two sectors. At this
point, the electrical potentials in the two sectors were equal and thus
identical to that of the known Daniell cell. They then removed the external
capacitor (C) and, accordingly, some weights until the needle had once again
stabilized at its zero point, that is, until the voltage across the plate again
corresponded to that of the Daniell cell. The difference in the amount of
charge between the two scenarios can be attributed to the different loads
exerted by the resting weight on the crystal. Because the voltage remained
constant, the difference in charge corresponded to the product of the known
voltage and difference in capacitance, which represents the capacitance of
the cylindrical capacitor ($\Delta Q = U \Delta C$). Thus, they were able to
directly derive the charge generated per unit change in weight or
force~\cite{Katzir2006}.

By determining only the differences in size, thenCuries avoided the need for
complex measurements of the system capacitance. Instead, they merely need to
know the capacitance of the known capacitor. They used ``a cylindrical
capacitor consisting of two (narrow) plate segments, with which errors caused
by the edges (extrémités) can be minimised'', and determined the capacitance
based on their dimensions using a method not further
described~\cite{Curie1882}. In developing a ``zero experiment'', in which they
held the needle of the electrometer at zero, the Curies reduced errors not only
in reading the needle deflection, but particularly in converting this
deflection into units of voltage. This made thendata analysis in their
experiment straightforward and required no complex mathematics. Maintaining a
low voltage (a few volts) during most parts of the experiment had the added
benefit of reducing electrical losses~\cite{Katzir2006}.

They determined the piezoelectric constant for quartz and tourmaline to be
$0.062$ (esu/kg) and $0.053$ (esu/kg) respectively. The unit esu represents the
electrostatic units and is based on the CGS system of mechanics. In units used
later, these values correspond to $6.3 \times 10^{-8}$ and $5.4 \times 10^{-8}$
(statcoulombs/dyn). These values are less than \SI{10}{\percent} below the
current values, presumably because of irregularities in the crystals examined.
The brothers further determine the coefficients. In 1882, they published
slightly adjusted results for quartz -- $0.063$ electrostatic units, which
brought them closer to the current values, but in 1889, they returned to
the original value of $0.062$~\cite{Curie1889}.

Jacques Curie continued to measure the coefficient for quartz using the same
``zero-experiment'' method, at least until the end of the first decade of the
20th century, and established a value of $6.9 \times 10^{-8}$, which agrees
with the current value~\cite{Cady1946}, \cite{Katzir2006}.

The brothers quickly discovered practical applications for empirical research.
They demonstrated that the newly developed device, which was intended to
determine the coefficients of the newly discovered phenomena, could also be
used to measure the electrical values. In the same publication in which they
presented the coefficient values, they explained how once the piezoelectric
coefficient of a crystal had been determined, its piezoelectric effect could be
used to determine capacitance, electrostatic force, and, in particular, charge.
Thus, the piezoelectric coefficient was used to derive further quantities. In
particular, it enabled the performance of ``zero experiments'' in various
electrical measurements using piezoelectric quartz to compensate for the
electrical effect under investigation. This represents the first step in a
series of measuring instruments based on the piezoelectric effect developed by
the brothers. Both used these devices in subsequnt studies. The development and
construction of physical instruments, particularly by experimental physicists,
are common among French physicists. The working methods of the Curie brothers
can be viewed as part of this French tradition~\cite{Katzir2006}.

The research achievements of Jacques and Pierre Curie, two young assistant
physicists who were 25 and 21 years old, respectively, at the time of their
first experiment, were remarkable. They discovered a new phenomenon and
conducted a thorough experimental analysis that yielded quantitative
relationships describing it. Their work shows a clear influence from research
on pyroelectricity, which they explicitly linked to piezoelectricity
in their very first publication on their findings and considerations. They
observed the effect in a targeted experiment designed to detect it in
axes known to be polarized by pyroelectricity and supplemented their earlier
experiments involving heating and cooling with further tests involving
pressure. Furthermore, three weeks after the publication of piezoelectric laws,
the Curies proposed a unified explanation for both pyro- and
piezoelectricity~\cite{Katzir2006}. While the Curies initially described the
piezoelectric effect purely phenomenologically, modern solid-state physics
allows for microscopic interpretation based on the crystal structure of the
material in question.

Using the example of low quartz, one of the key crystals studied by the
Curies, the following section demonstrates how the macroscopically observed
electrical polarization results from the asymmetric displacement of
silicon $\left(\mathrm{Si}^{4+}\right)$ and oxygen ions
$\left(\mathrm{O}^{2-}\right)$ within the $[\mathrm{SiO_4}]^{4-}$ tetrahedral
network. The role of crystal symmetry is crucial here: the symmetry conditions
empirically formulated by the Curies in 1882 correspond exactly to those
constraints that can also be derived from the microscopic structure and its
transformation properties.

\section{The physical principle behind piezoelectricity}
\label{sec:Physik}

\subsection{Crystallographic fundamentals: Polar axes in
low quartz}

\begin{figure}[t]
\centering
\includegraphics[width=\linewidth]
{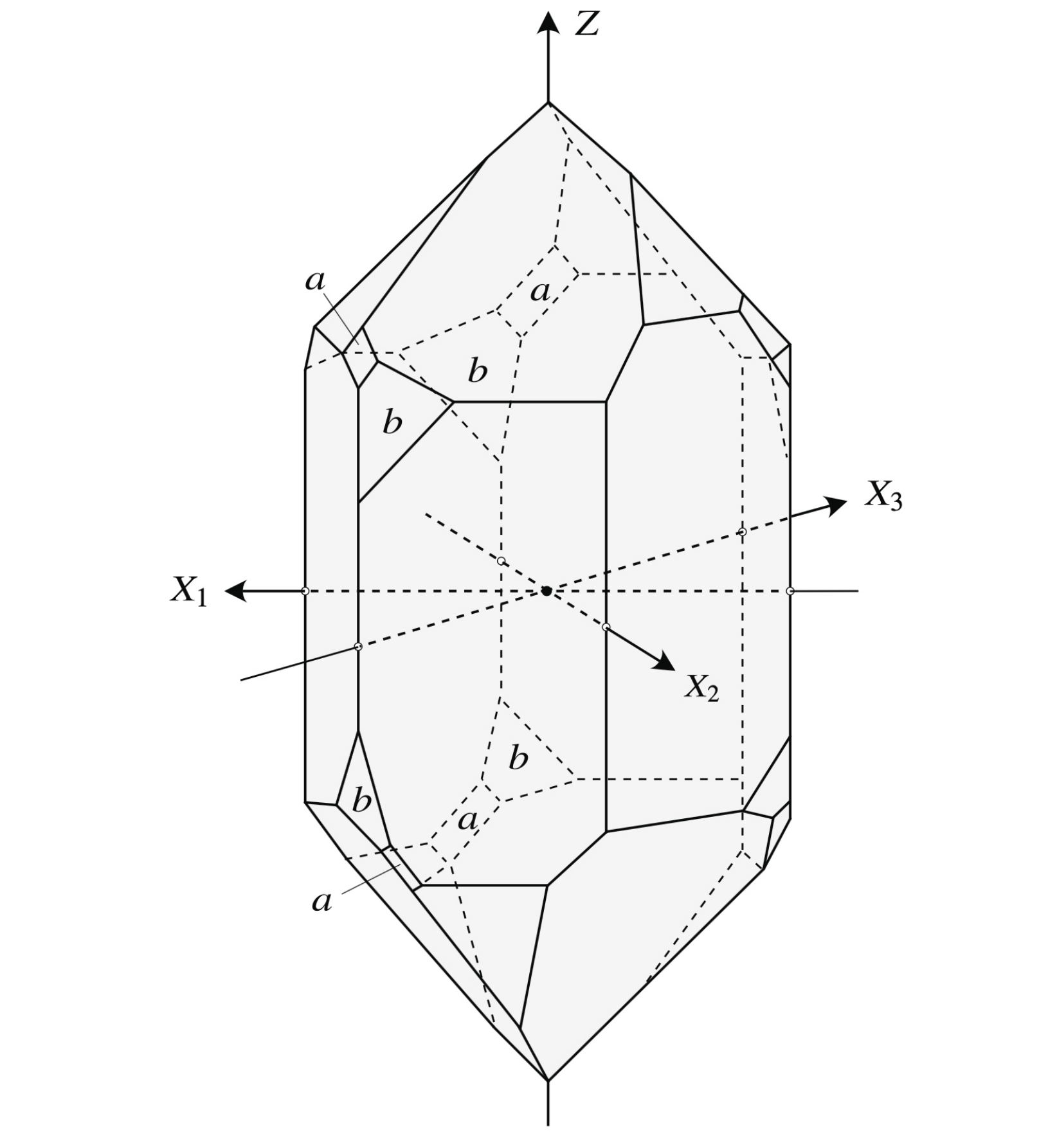}
\caption{Crystal form of low quartz with its
corresponding crystal axes (reproduced from~\cite{Raith2006}).}
\label{fig:Tiefquarz}
\end{figure}

A common characteristic of all piezoelectric crystals is the presence of one or
more polar axes. In crystallography, the polar axis is characterized by the
fact that its front and rear ends are not equivalent, that is, a rotation of
\SI{180}{\degree} about an axis perpendicular to the polar axis does not bring
the relevant part of the crystal back into alignment with its original
position~\cite{Raith2006}. This can be illustrated at the macroscopic level
using the crystal form of quartz shown in Figure~\ref{fig:Tiefquarz}, often
referred to as low quartz~\cite{Ballas2025a}.

There are three polar axes, labelled $X_1$, $X_2$ and $X_3$ (see
Figure~\ref{fig:Tiefquarz}). These axes connect two opposite edges of the
hexagonal prism. The opposite edges are, however, not identical, as can be seen
from the fact that, for example, on the rear edge, which is assigned to the
$X_2$ axis, the small faces are labelled $a$ and $b$, while these are absent on
the opposite edge. This implies that rotation at \SI{180}{\degree} about the
axis labelled $Z$ does not return the quartz crystal to its original position.
The $Z$-axis represents the crystallographic principal axis (often referred
to as the optical axis) and is non-polar. This is because a \SI{180}{\degree}
rotation about one of the $X$-axes brings the quartz crystal back to alignment
with its initial position~\cite{Cady1946}, \cite{Raith2006}.

\begin{figure}[t]
\centering
\includegraphics[width=\linewidth]
{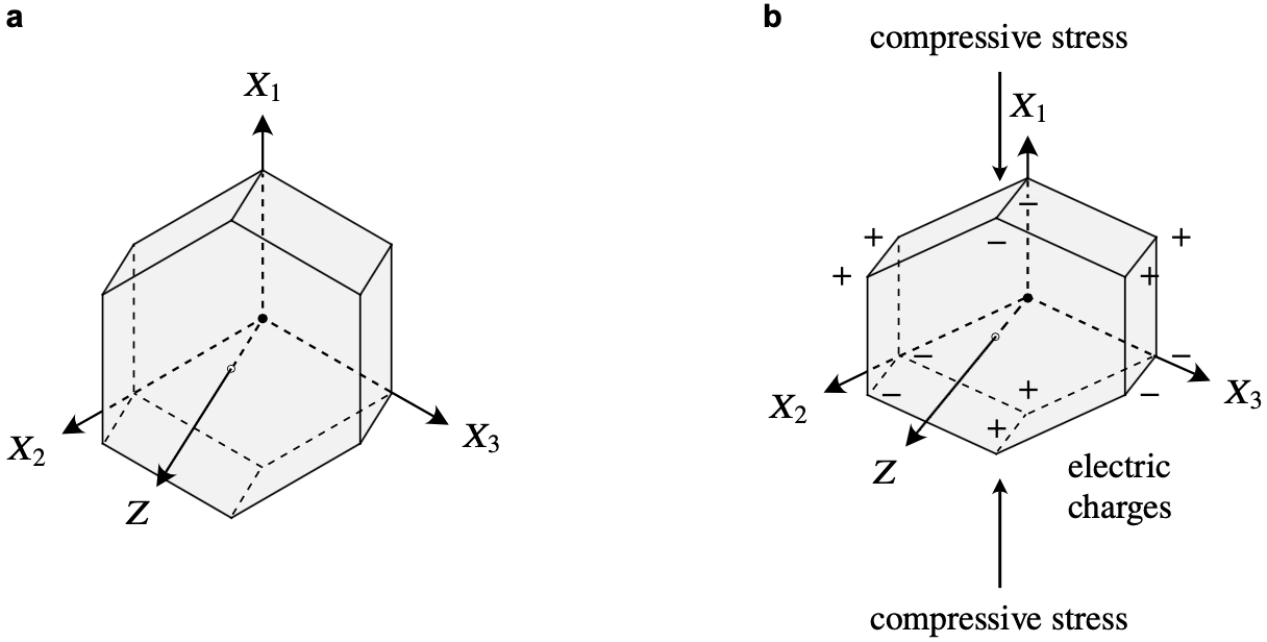}
\caption{Generation of piezoelectricity
in a six-sided quartz prism. \textbf{a} The quartz prism is initially in an
unloaded state. \textbf{b} When a mechanical compressive load is applied,
opposite but equal electrical charges appear at the edges of the quartz prism,
i.e. piezoelectricity is generated~\cite{Ballas2025a}.}
\label{fig:Quarzprisma}
\end{figure}

Figure~\ref{fig:Quarzprisma}a shows a six-sided prism, which can be thought of
as a section of the quartz crystal shown in Figure~\ref{fig:Tiefquarz}, cut
parallel to the plane of the polar axes $X_1$, $X_2$ and $X_3$. This prism was
subjected to a mechanical compressive load along the $X_1$-axis (see
Figure~\ref{fig:Quarzprisma}b).
It can be observed that when a six-sided quartz prism is subjected to
mechanical stress, equal but opposite electrical charges are generated at the
ends of the respective polar axes, which signifies the generation of 
piezoelectricity. In this context, it is not strictly necessary for mechanical
stress to be applied exactly in the direction of the polar axis. A compression
component in the direction of the polar axis is sufficient for this.

\begin{figure}[b]
\centering
\includegraphics[width=\linewidth]
{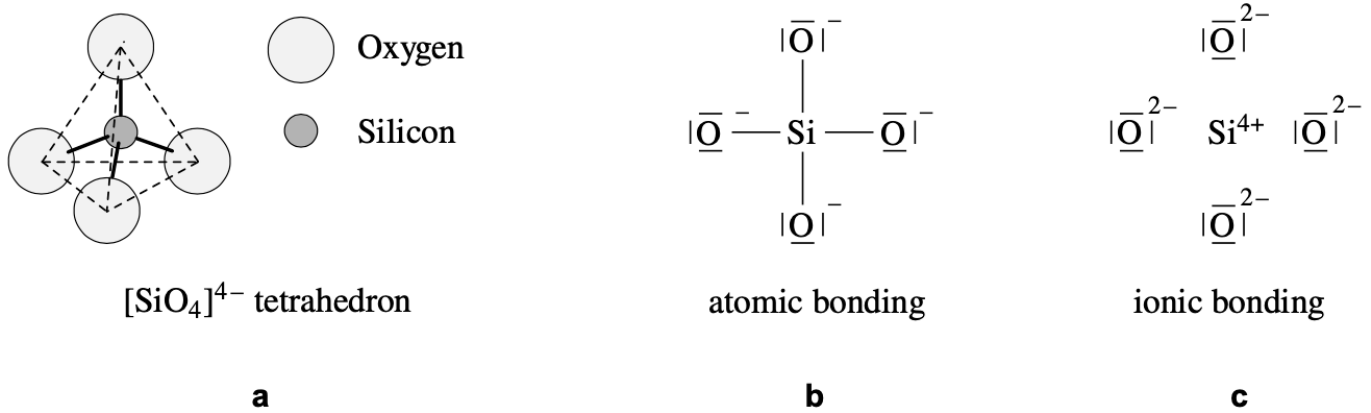}
\caption{Structural coordination polyhedron of low quartz and its 
mesomeric limit structures (the valence electron pairs are indicated by lines).
\textbf{a} $[\mathrm{SiO_4}]^{4-}$ tetrahedron (the silicon atom is surrounded
by four oxygen atoms). \textbf{b} Purely covalent bond (also known as an atomic
bond, homopolar bond or electron pair bond). \textbf{c} Purely ionic bond (also
known as an ionic bond, heteropolar bond or electrovalent
bond)~\cite{Ballas2025a}.}
\label{fig:Koordinationspolyeder}
\end{figure}

A more detailed investigation of the structural composition of rock crystals
and their molecular chemical bonds provides a physical basis for the
piezoelectric phenomenon, which is evident in quartz prisms~\cite{Raith2006}.
low quartz was the most significant naturally occurring form of
silicon dioxide ($\mathrm{SiO_2}$). At the molecular level, it can be regarded
as a network of continuously interconnected $[\mathrm{SiO_4}]^{4-}$
tetrahedra~\cite{Davila2008}, \cite{Raith2006}. In silicates to which
low quartz belongs, $[\mathrm{SiO_4}]^{4-}$ tetrahedra form the central
structure-defining coordination polyhedra. Here, the silicon atom is
surrounded by four oxygen atoms (see Figure~\ref{fig:Koordinationspolyeder}a).

With regard to the $\mathrm{Si}-\mathrm{O}$ bond and the associated bonding
relationships within the $[\mathrm{SiO_4}]^{4-}$ tetrahedron of 
low quartz, this is neither purely covalent (see
Figure~\ref{fig:Koordinationspolyeder}b) nor purely ionic (see
Figure~\ref{fig:Koordinationspolyeder}c). The actual bonding conditions lie
somewhere between those in~\cite{Ballas2025a}. Analysis of the
electronegativities of the bond partners involved and the resulting
electronegativity difference provides an indication of the dominant bonding
characteristics~\cite{Mortimer2015}. For the $\mathrm{Si}-\mathrm{O}$ bond in 
low quartz, the electronegativity of silicon ($\chi_{\mathrm{Si}} =
1.8$) and oxygen ($\chi_{\mathrm{O}} = 3.5$) results in a difference of 
$\Delta\chi_{\mathrm{Si}}=1.7$. Given this value, the $\mathrm{Si}-\mathrm{O}$
bond is assumed to have a partially ionic character of approximately
\SI{50}{\percent}~\cite{Mortimer2015}.

\begin{figure}[b]
\centering
\includegraphics[width=\linewidth]
{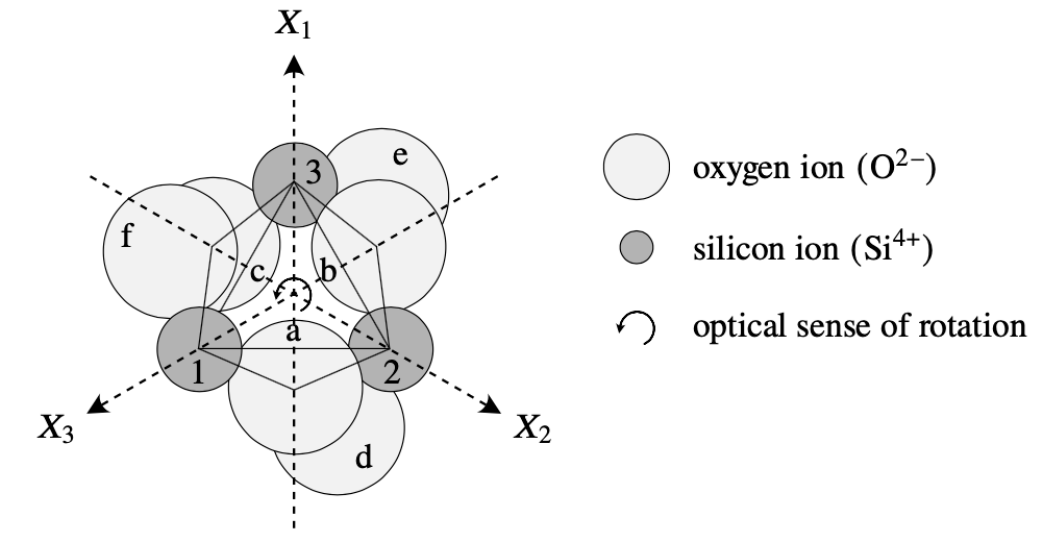}
\caption{Unit cell of low quartz with the corresponding
polar axes $X_1$, $X_2$ and $X_3$ (the crystallographic principal axis $Z$
points perpendicular into the plane of the diagram) (adapted
from~\cite{Gobrecht1971}.}
\label{fig:Strukturzelle}
\end{figure}

Assuming a purely ionic bond within the $[\mathrm{SiO_4}]^{4-}$ tetrahedron,
the structure of the unit cell of the crystalline quartz can be clearly
illustrated. Owing to the characteristic spatial interlinking of the
$[\mathrm{SiO_4}]^{4-}$ tetrahedra in low quartz, a six-sided unit cell
results in a first approximation (see Figure~\ref{fig:Strukturzelle}). It
consists of three formula units $\mathrm{SiO}_2$~\cite{Ballas2025a}.
The assumption of purely ionic bonds implies that the silicon ion (cation)
within the unit cell of the low quartz occupies a smaller volume than
the oxygen ion (anion). These differences are illustrated in
Figure~\ref{fig:Strukturzelle} by the different representations of the sizes of
silicon and oxygen ions~\cite{Ballas2025a}. The main crystallographic axis
$Z$ of the low quartz lies perpendicular to the plane spanned by the 
polar axes $X_1$, $X_2$ and $X_3$, and points into the drawing plane
(see Figure~\ref{fig:Strukturzelle}). Relative to this plane, silicon ion
$1$ lies above silicon ion $2$, which, in turn, lies above silicon ion $3$. The
arrangement of the oxygen ions, each offset by \SI{60}{\degree}
$\mathrm{a}$, $\mathrm{b}$ and $\mathrm{c}$, followed accordingly. From this,
it is clear that the positions of the silicon and oxygen ions extend spirally
in an anti-clockwise direction into the plane of the
drawing~\cite{Gobrecht1971}. The direction of rotation of the spiral in
low quartz reflects the optical rotation of the crystal structure, which
means that the plane of the linearly polarized light is rotated anticlockwise. 
Figure~\ref{fig:Strukturzelle} shows the unit cell of a left-handed quartz
crystal~\cite{Gobrecht1971}. The unit cell of low quartz appears
externally electrically neutral, as each of the three silicon ions has four
positive unit charges and each of the six oxygen ions has two negative unit
charges, whereby all charges cancel each other out (see
Figure~\ref{fig:Strukturzelle})~\cite{Raith2006}.

\begin{figure}[b]
\centering
\includegraphics[width=\linewidth]
{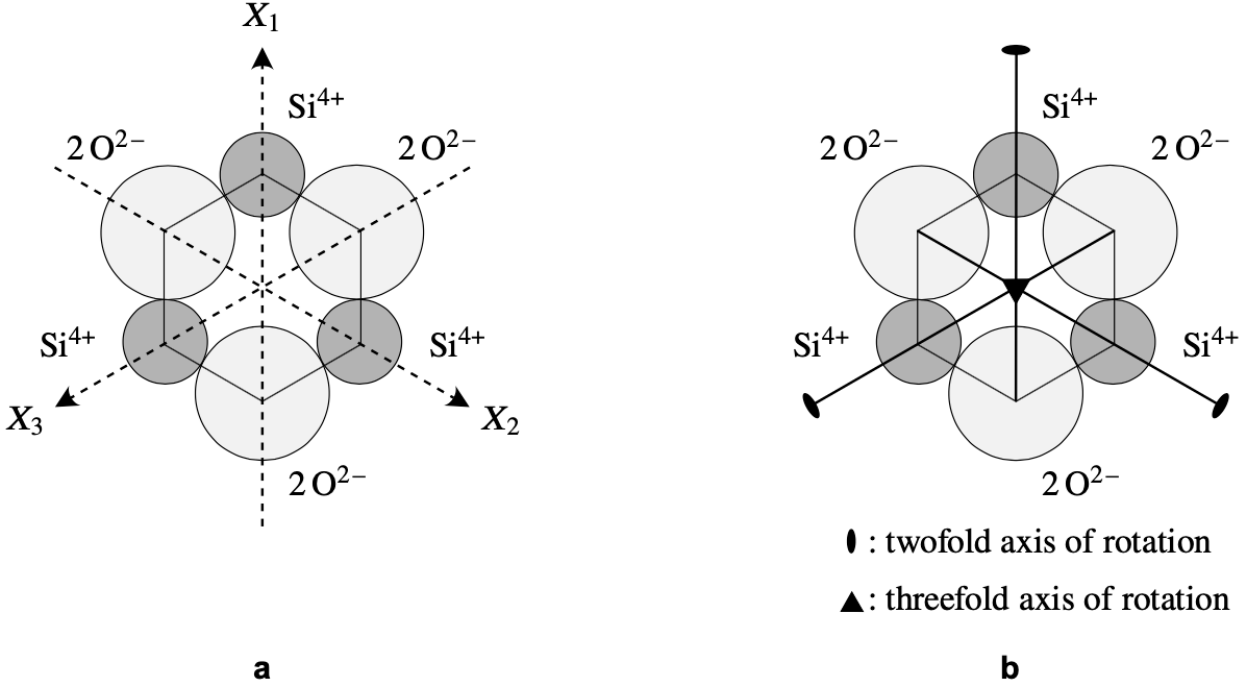}
\caption{Simplified form of the unit cell of low quartz with the
associated polar axes $X_1$, $X_2$ and $X_3$. \textbf{a} Arrangement of the
$\mathrm{Si}^{4+}$ and $\mathrm{O}^{2-}$ ions in a regular hexagon. \textbf{b}
Two-fold and three-fold axes of rotation as symmetry
elements~\cite{Ballas2025a}.}
\label{fig:vereinfachte Strukturzelle}
\end{figure}

For simplicity, it is assumed that the oxygen ions $\mathrm{d}$,
$\mathrm{e}$, and $\mathrm{f}$ are slightly displaced so that they coincide
with the corresponding oxygen ions $\mathrm{a}$, $\mathrm{b}$ and $\mathrm{c}$
along the main crystallographic axis. This resulted in a simplified
representation of the unit cell of low quartz (see
Figure~\ref{fig:vereinfachte Strukturzelle}). This shows a regular hexagonal
unit cell, at the corners of which negatively charged $\mathrm{O}^{2-}$ and
positively charged $\mathrm{Si}^{4+}$ ions are arranged~\cite{Raith2006}.
The polar axes $X_1$, $X_2$ and $X_3$ each act as two-fold rotation axes,
whereas the crystallographic principal axis $Z$, which is perpendicular to the
plane of the diagram, represents a threefold axis of rotation. In
crystallography, the axis of rotation is referred to as the $n$-fold axis of
rotation if a crystal lattice, after a rotation of $\SI{360}{\degree}/n$,
achieves a configuration that no longer differs from the initial form of the
crystal lattice~\cite{Paufler1975}, \cite{Borchardt-Ott2012}. For example,
rotation of the unit cell of deep quartz, as shown in
Figure~\ref{fig:vereinfachte Strukturzelle}b, about the polar axis $X_1$ by
\SI{180}{\degree} results in a configuration that corresponds to the initial
situation~\cite{Ballas2025a}. Similarly, the threefold nature of the main
crystallographic axis $Z$ can be justified. A rotation of \SI{180}{\degree}
constitutes a symmetry operation, whereby the two-fold rotation axis
($\SI{360}{\degree}/2=\SI{180}{\degree}$) forms them corresponding symmetry
element. If this symmetry operation is applied twice in succession, all ions in
the unit cell return to their original positions~\cite{Borchardt-Ott2012}.

\subsection{Direct piezoelectric effect}

\subsubsection{Longitudinal effect}

\begin{figure}[b]
\centering
\includegraphics[width=\linewidth]
{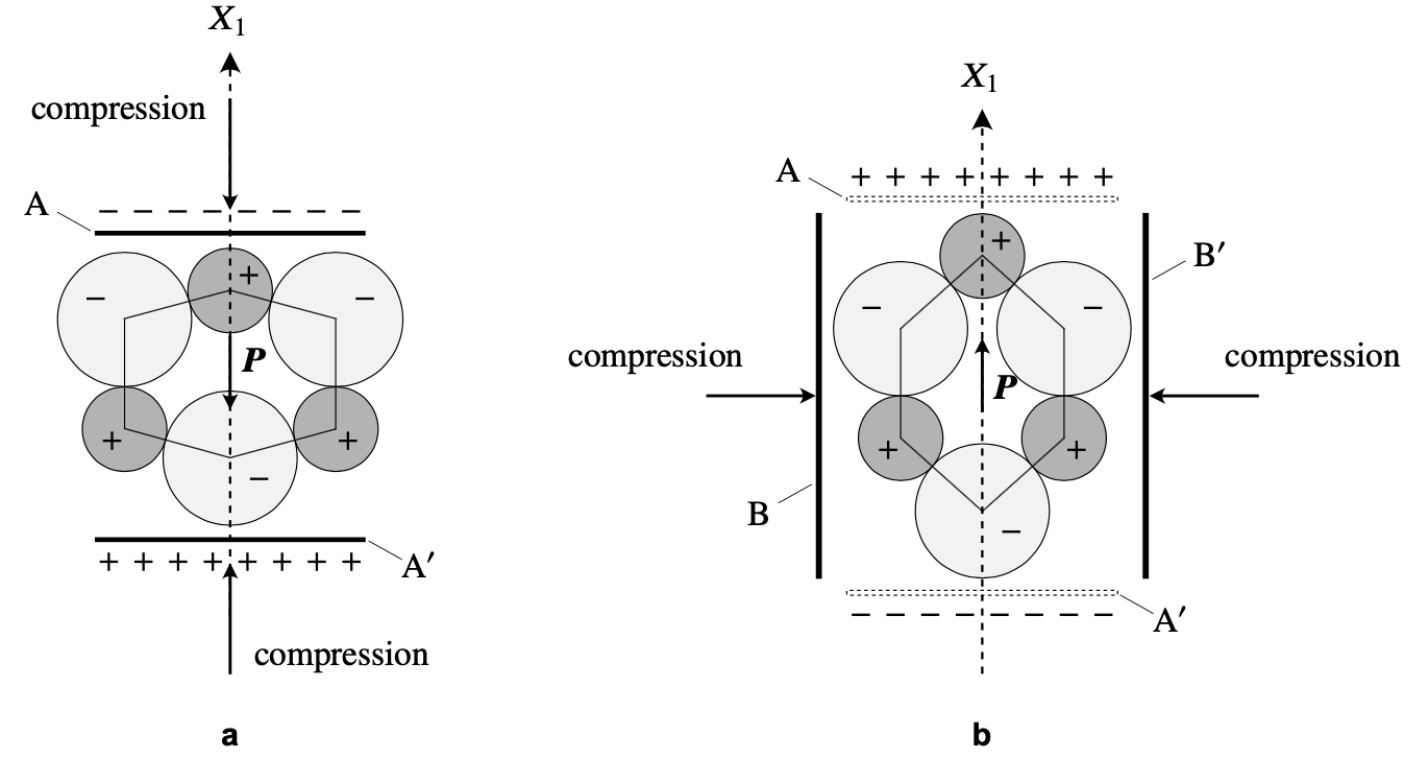}
\caption{Direct piezoelectric effect within the structural cell of
low quartz. \textbf{a} Longitudinal piezoelectric effect (the
vector of the electric polarization $\boldsymbol{P}$ runs parallel to the
direction of the mechanical compressive stress). \textbf{b} Transverse
piezoelectric effect (the vector of the electric polarization $\boldsymbol{P}$
is perpendicular to the direction of the mechanical
compressive stress)~\cite{Ballas2025a}.}
\label{fig:direkter Effekt}
\end{figure}

If the simplified unit cell of low quartz is mechanically compressed
along the polar axis $X_1$ (see Figure~\ref{fig:direkter Effekt}), the upper
silicon ion shifts between the two upper oxygen ions, whereas the lower oxygen
ion slides between the two lower silicon ions~\cite{Raith2006}. The relative
displacement of the positively and negatively charged ions within the unit cell
(see Figure~\ref{fig:direkter Effekt}) generates an electrical polarization 
$\boldsymbol{P}$ along the polar axis $X_1$, which coincides with the direction
of the mechanical compressive stress. This induces negative charges at the
upper electrode $\mathrm{A}$ and positive charges at the lower electrode
$\mathrm{A}^\prime$, which cause an external electrical polarization voltage.
This effect is referred to as the direct longitudinal piezoelectric effect and
explains, at the macroscopic level, the piezoelectric phenomenon along polar
axis $X_1$ of the pressure-loaded quartz prism shown in
Figure~\ref{fig:Quarzprisma}b~\cite{Ballas2025a}.

\subsubsection{Transversal effect}

On the other hand, if the unit cell is mechanically compressed perpendicular
to the polar axis $X_1$, the silicon and oxygen ions on the left and right of
the structural cell shift uniformly inward, causing their charges to cancel
each other out. Consequently, no charges are
induced at the outer electrodes $\mathrm{B}$ and $\mathrm{B}^\prime$
(see Figure~\ref{fig:direkter Effekt}b)~\cite{Raith2006}. The upper silicon and
lower oxygen ions of the deep quartz unit cell, on the other hand, are
displaced outwards, which generates an electric polarization $\boldsymbol{P}$
along the polar axis $X_1$, but rotated by \SI{180}{\degree} and which is
perpendicular to the direction of the mechanical stress. Consequently, positive
charges arise at the upper electrode $\mathrm{A}$ and negative charges at the
lower electrode $\mathrm{A}^\prime$, which induce an external polarization
voltage with the opposite sign. This effect is referred to as the direct
transverse piezoelectric effect~\cite{Raith2006}.

\subsection{Electric dipole and electrical polarization}

\begin{figure}[h]
\centering
\includegraphics[scale=0.5]
{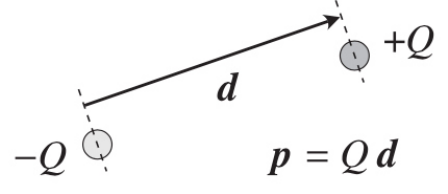}
\caption{Definition of the electric dipole moment. The arrangement of two
opposite charges of equal magnitude at a small distance from one another
is referred to as an electric dipole; the electric dipole moment
$\boldsymbol{p}$, defined as the product of the distance vector
$\boldsymbol{d}$ and the magnitude $Q$ of the charges, is a measure of the
electric strength of a dipole~\cite{Ballas2025a}.}
\label{fig:Dipolmoment}
\end{figure}

An electric dipole consists of two equal but opposite electric charges $\pm Q$,
which are separated by a distance $d$ (see Figure~\ref{fig:Dipolmoment}).
Because the opposite charges cancel each other out, the dipole remains
electrically neutral overall. This is described by the electric dipole moment, 
$\boldsymbol{p}$.

\begin{figure}[b]
\centering
\includegraphics[scale=0.5]
{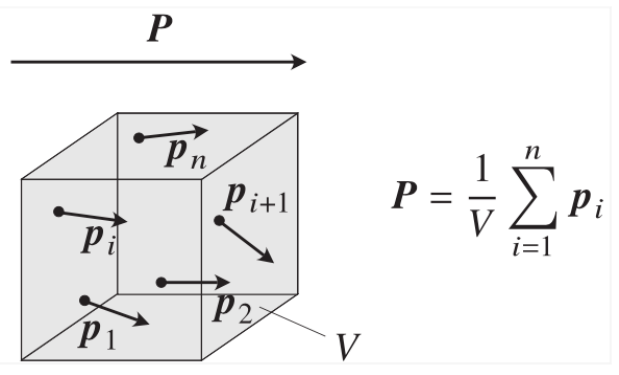}
\caption{The electric polarization $\boldsymbol{P}$ is defined as
the volume-normalised vector sum of the electric dipole moments
$\boldsymbol{p}_i$ of the unit cells~\cite{Ballas2025a}.}
\label{fig:Polarisation}
\end{figure}

If a system consisting of $n$ electrically active dipoles is located within
volume $V$, the vector sum of all individual dipole moments $\boldsymbol{p}_i$
gives rise to the total dipole moment $\boldsymbol{p}_{\Sigma}$:
\begin{equation}
	\boldsymbol{p}_{\Sigma}=
	\sum_{i=1}^{n}\boldsymbol{p}_i
	\label{eq:Gesamtdipolmoment}
\end{equation}
The ratio of the total dipole moment $\boldsymbol{p}_{\Sigma}$ to volume $V$
determines the electric polarization $\boldsymbol{P}$ (see
Figure~\ref{fig:Polarisation})~\cite{Ballas2025a}.
\begin{equation}
	\boldsymbol{P}=
	\dfrac{1}{V}\sum_{i}^{n}\boldsymbol{p}_i
	\label{eq:Polarisation}
\end{equation}

\subsection{Reciprocal piezoelectric effect}

Both variants of the direct piezoelectric effect are reversible, which means
that under the influence of appropriately aligned electric fields the unit cell
of low quartz contracts or expands. This process is referred to as the
reciprocal piezoelectric effect~\cite{Raith2006}, \cite{Hankel1881},
\cite{Voigt1966}.

\begin{figure}[b]
\centering
\includegraphics[width=\linewidth]
{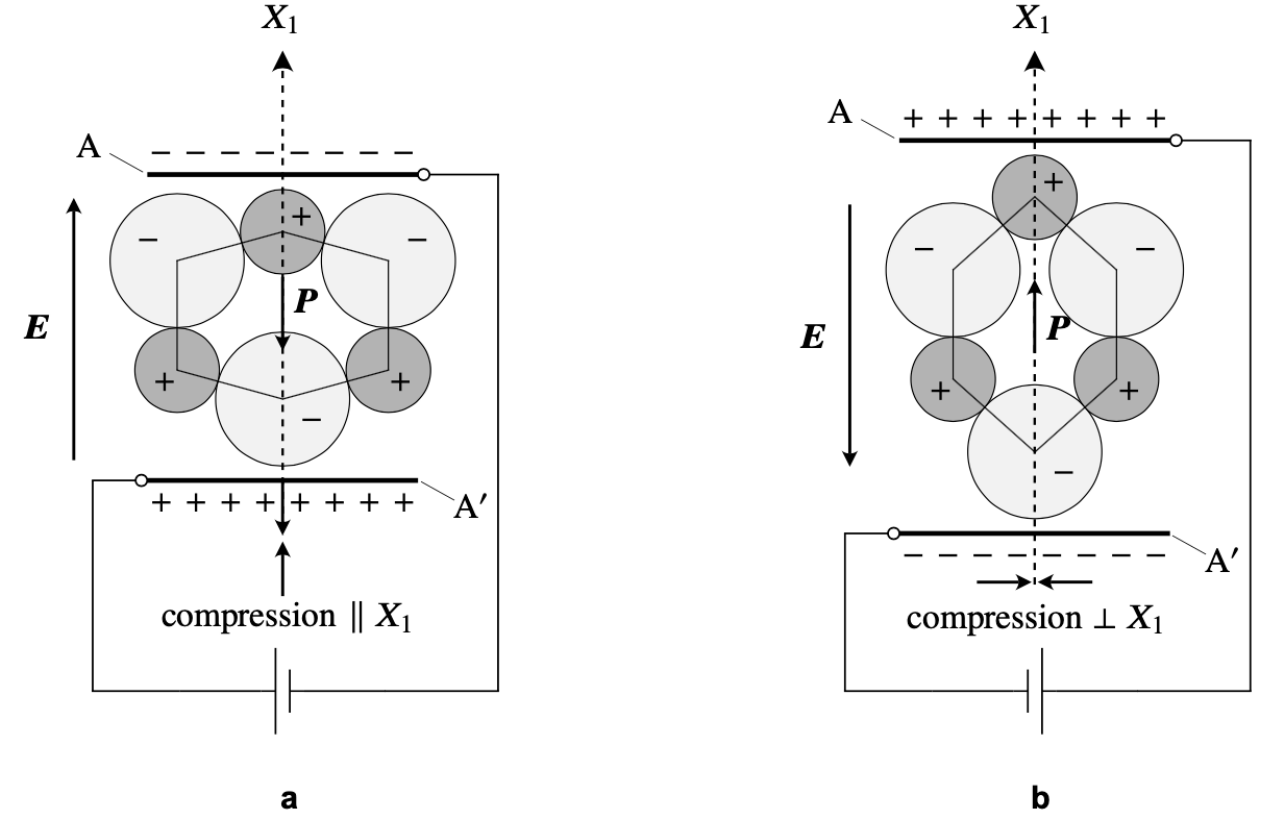}
\caption{Reciprocal piezoelectric effect within the unit cell of
low quartz. \textbf{a} Longitudinal piezoelectric effect (the
vector of electric polarization $\boldsymbol{P}$ runs parallel to the
compression direction of the unit cell). \textbf{b} Transverse piezoelectric
effect (the vector of electric polarization $\boldsymbol{P}$ runs perpendicular
to the compression direction of the unit cell)~\cite{Ballas2025a}.}
\label{fig:reziproker Effekt}
\end{figure}

If an external DC voltage is applied between the upper electrode $\mathrm{A}$
and the lower electrode $\mathrm{A}^\prime$, such that the upper electrode
$\mathrm{A}$ is negatively charged and the lower electrode $\mathrm{A}^\prime$
is positively charged, an electric field with electric field strength 
$\boldsymbol{E}$ is generated. The electric field strength $\boldsymbol{E}$ is
similar to the electric polarization $\boldsymbol{P}$ resulting from
deformation, a vector quantity, and, by definition, points from positive to
negative charge (see Figure~\ref{fig:reziproker Effekt}a)~\cite{Ballas2025a}.

When the voltage source is connected to the electrodes, as shown in
Figure~\ref{fig:reziproker Effekt}a, this causes the two upper oxygen ions
to be partially attracted towards the lower electrode $\mathrm{A}^\prime$ and
the two lower silicon ions to be partially attracted towards the upper
electrode $\mathrm{A}$. As a result, the structural cell undergoes compression
along polar axis $X_1$. Mutual displacement of positively and negatively
charged ions additionally generate an electric polarization $\boldsymbol{P}$
along the polar axis $X_1$, which coincides with the direction of compression
of the structural cell. This process is referred to as the reciprocal
longitudinal piezoelectric effect.

On the other hand, if the voltage source is connected in such a way that the
upper electrode $\mathrm{A}$ becomes positive and the lower electrode
$\mathrm{A}^\prime$ becomes negative, the direction of the electric field
strength $\boldsymbol{E}$ reverses by \SI{180}{\degree}. As a result, the two
upper oxygen ions are partially drawn towards the upper electrode $\mathrm{A}$
and the two lower silicon ions are partially drawn towards the lower electrode
$\mathrm{A}^\prime$. Simultaneously, the silicon and oxygen ions on the left
and right of the unit cell shift uniformly inward (see
Figure~\ref{fig:reziproker Effekt}b). The displacement of the positively and
negatively charged ions again generates an electrical polarization
$\boldsymbol{P}$ along the polar axis $X_1$, but rotated by \SI{180}{\degree}
and perpendicular to the compression direction of the unit cell. This effect is
referred to as reciprocal transverse piezoelectric effect.

\subsection{Important piezoelectric crystals and applications}

The crystalline form low quartz, which is formed by quartz and remains
stable up to \SI{573}{\degreeCelsius}, is regarded as the most significant
example of a piezoelectric crystal. It is primarily used as a quartz crystal,
an electronic component in oscillator circuits that generates electrical
oscillations at a fixed frequency~\cite{Ballas2025a}.
Other important piezoelectric crystals include lithium niobate
($\mathrm{LiNbO}_3$), gallium orthophosphate ($\mathrm{GaPO}_4$) and
langasite ($\mathrm{La_3Ga_5SiO_{14}}$). Unlike quartz, these do not occur
naturally, but are produced exclusively synthetically. On the one hand, they
are characterized by significantly higher piezoelectric constants compared to
quartz; on the other hand, their phase transition occurs at significantly
higher temperatures
($\SI{970}{}$\,--$\SI{1{,}470}{\degreeCelsius}$)~\cite{Heywang2008},
\cite{Jiang2013}. Their applications include piezoelectric SAW filters (surface
acoustic wave) for signal detection and processing in modern telecommunications
systems, SAW sensors for the wireless measurement of temperature,
force and torque on stationary or moving components, and piezoelectric
pressure transducers for high-temperature applications~\cite{Tichy2010}.

\section{Conclusion}

The discovery of piezoelectricity by Jacques and Pierre Curie in 1880 marked a
milestone in solid-state physics. Through the skillful combination of
crystallographic considerations regarding pyroelectricity and precise
quantitative measurements, the brothers succeeded within a very short time, not
only to demonstrate a previously unknown phenomenon, but also to describe it
mathematically through initial empirical laws. Their pioneering work laid the
foundation for understanding that the mechanical deformation and electrical
polarization in anisotropic materials are inextricably linked.

Physical investigations have shown that the presence of polar axes in
hemihedral crystals is a fundamental prerequisite for this effect. Using
low quartz as an example, it was demonstrated microscopically
that the displacement of the centers of charge within the
$[\mathrm{SiO_4}]^{4-}$ tetrahedra leads to the formation of a total dipole
moment. Notably, the reversibility of the effect is worth noting. While the
direct piezoelectric effect forms the basis for sensor technology, the
reciprocal effect enables precise actuation as well as the generation of stable
electrical oscillations. 

Currently, piezoelectricity has become indispensable in modern technology.
Quartz crystals form the heart of almost every electronic timing and clocking
system. Furthermore, synthetic high-temperature materials such as lithium
niobate ($\mathrm{LiNbO}_3$) or langasite ($\mathrm{La_3Ga_5SiO_{14}}$) have
expanded the range of applications to include extreme environments and wireless
sensor technology (SAW technology)~\cite{Heywang2008}, \cite{Jiang2013}. 

The outlook for piezoelectric research is promising: the development of
lead-free ceramics to meet environmental standards~\cite{Rodel2009}, as
well as research into piezoelectric nanostructures for ``energy
harvesting''~\cite{Anton2007}, \cite{Wang2006} promise innovative solutions for
powering self-sufficient microsystems. Thus,c the research discipline founded
by
the Curies remains a key driver of technological progress even after almost 150
years.

\backmatter

\bibliography{Piezoelectricity}% common bib file
%% if required, the content of .bbl file can be included here once bbl is
%generated
%\input Pyroelectricity.bbl

\end{document}